# Exploring the Self-enhanced Mechanism of Interactive Advertising Phenomenon---Based on the Research of Three Cases

Ren jian[1] and Ding Wanxing[2]

[1]College of Communication and Art Design, University of Shanghai for Science and Technology, Shanghai, China
moomsincere@126.com

[2]College of Communication and Art Design, University of Shanghai for Science and Technology, Shanghai, China
270258632@qq.com

## *Abstract*

*Under the background of the new media era with the rapid development of interactive advertising, this paper used case study method based on the summary of the research of the communication effect of interactive advertising from both domestic and foreign academia. This paper divided interactive advertising into three types to examine ---- interactive ads on official website, interactive ads based on SNS and interactive ads based on mobile media. Furthermore, this paper induced and summarized a self-enhanced dissemination mechanism of the interactive advertising, including three parts which are micro level, meso level and macro level mechanism, micro level embodies core interaction, inner interaction and outer interaction which reveal the whole process of interact with contents, with people and with computer, and the communication approach and spread speed shown in meso level which is self-fission-type spread, finally in macro level the communication effect of IA achieved the spiral increasing. In a word, this article enriches research procedure of the interactive advertising communication effects.*

## 1. INTRODUCTION

The increasing research of Interactive advertising (IA) is related to the web.2.0, and the interactivity is considered as the key feature compared to traditional advertisement as well as the main factor of successful online advertising. While the former researches were mostly focus on one specific type of interactive ads, the mechanism of ads' effect is hard to apply into other types of interactive ads, so in this article, the author will analyze three cases related to interactive advertising , namely , official website interactive ads, SNS interactive ads, and mobile media interactive ads, from a communication perspective, based on a personal view that has been informed by a review of coverage by a number of related research , both qualitative and quantitative , domestic and foreign , to explore whether a comprehensive mechanism is exist or not.

## 2. LITERATURE REVIEW

### 2.1. Definition of interactivity (connotation and denotation)

Along with the continuous development of the network and the rise of its' study, interactive advertising came into the view of research scholars, and great attention has been put on to the key factor of interactive ads, namely, *interactivity*. Interactivity is recognized as the key features of new media which is different from traditional media (Chung & Zhao, 2004) as well as a key factor of the success of online advertising (Kim & McMillan2008). Therefore, the definition of research on the interaction interactive advertising will naturally become the focus of academic attention.

Rafaeli(1988)[1] introduced interactivity's concept firstly, Rafaeli's definition refers to "interactivity as an expression of the extent that in a given series of communication exchanges, any third (or later) transmission ( or message ) is related to even earlier transmission." This definition created a new era of interactivity which became a basic communication concept, that is to say, interpersonal communication and mass communication has been connected from the perspective of interactivity. Then researchers have continued to expand on the basic definition. A great deal of research on interactivity has focused on the bilateral involved in information exchange, whereby scholars have identified and classified the characteristics of interactivity which include bilateral sides, which are same with the Rafaeli's definition. Blattberg and Deighton(1991)[2] refers interactivity as "the facility for individuals and organizations to communicate directly with each other despite time or distance". Steuer's(1992)[3] definition refers to "the degree to which users of a medium can influence the form or content of the mediated environment". Compared with the former definition, not only did Steuer emphasize the interaction between bilateral sides, also he lay emphasis on users' initiative power to modify and control media. Deighton (1996) [4] continued to extend the definition of interactivity, he pointed out "interactivity requires two features of communication—the ability to communicate with an individual and the ability to gather and remember that individual's response". This definition lays emphasis on the communication between organization and the individuals rather than organization and public.

Hoffman and Novak (1996)[5] combined Rafaeli's definition and Steuer's together which refers to "interactivity in hypermedia CMEs, like Web sites on the Internet , can happen with the medium(i.e..machine interactivity)in addition to through the medium (i.e..person interactivity)". Thus the definition of interactivity has been largely shaped, other definition pointed out by scholars are mostly based on it which is just emphasis different aspects. Roman et al (1996) considered interactivity is the interaction between the sender and receiver during communication process which focused on the bilateral interaction. Ha and James (1998)[6]defined interactivity as "the extent to which the communicator and the audience respond to, or are willing to facilitate, each other's communication goal", which emphasized the degree of interaction between bilateral sides. Ghose and Dou [7](1998) pointed out that " interactivity makes it possible to address the individual again and take into account his/her previous response", which lay emphasis on the degree of attention of the audience's feedback. Rice(1999)suggested that interactivity is the degree of the involver hold to control the conversation and exchange the communication role, Rice notion of interactivity provides a broad conceptualization of the power possessed by involvers in new media era of high degree of active participation and control ability.

Historically speaking, Grace is often considered the great master of definition of interactivity owning to the comprehensive summary of the former researches. Grace et al (2006) summarized and classified four shortcomings inherent by former researches, that is, the definition of IA was lack of the compatibility and globalization characterized, lack of the basic introduction associated with the dimensions of the interactivity, lack of clear definition of the aspect of nonverbal communication in the interactivity research, and the isolation of the related research areas of interactive advertising. Thereby, they suggested interactivity is the extent to which an actor involved in a communication episode perceives the communication to be reciprocal, responsive, speedy and characterized by the use of nonverbal information, which is a strong and wide range applicable definition embedded four facets, that is, reciprocity, responsiveness, the speedy of response and nonverbal information.

Finally, in summary , the academia's research on IA undergoing a deep and meaningful change which is turning from absolute way to relative way, from single values to dialectical values, therefore, which provide a more scientific and rigorous theoretical foundation to explore the mechanism of IA effect future in the future.

## 2.2. Related communication theory

The biggest creation of this article is the assumption of the self-enhanced model which is also inspired by some classical theory of communication effect, that is, Dance theory and the Spiral of Silence, which it is also the creation of this article by combine the IA and communication theory together.

Dance has examined and pointed out helical communication model in *Human Communication Theory* in 1967, he suggested that at any and all times, the helix gives geometrical testimony to the concept that communication while moving forward is at the same moment coming back upon itself and being affected by Its past behavior, for the coming curve of the helix is fundamentally affected by the curve from which it emerges. Yet, even though slowly, the helix can gradually free itself from its lower-level distortions. The communication process, like the helix, is constantly moving forward and yet is always to some degree dependent upon the past, which informs the present and the future. The helical communication model offers a flexible communication process.

Compared to traditional theory like 5W-model, Dance's helical communication model is an important turning point during the history of communication theory by laying emphasis on audiences' creativity and initiative ability.

Similarly, the Spiral of Silence proposed by Noelle Neumann emphasized a helix communication model too. She pointed out that strong public opinion tends to silent the weak public opinion, gradually strong public opinion get stronger and stronger, weak public opinion tends to be weaker and weaker, then a spiral of silence formatted, which will help to built a mainstream views. The spiral model proposed by the two theories and the audience-centered philosophy contained in these two models tends to adapt to the research in new media era, which also provide applicable theoretical implications to current IA research focus on audiences' interaction and feedback.

## 2.3. The internal dissemination mechanism of interactive advertising

As we all know, need generate consumption. Apply it into communication area it becomes to the audience's psychological need generate the dissemination of advertising. From shopping motivation to information gathering motivation, from surfing Internet for entertainment to communicating with others or seeking social escaping, there are approximately more than 100 different types of Internet use motivations among these articles.

Recently, examining the effect of interactive advertising through the perspective of the audience's psychological need becomes one of the main research directions. Hanjon et al (2005)[18] list four Internet use motivations: information , convenience , entertainment and social interaction. And they also find that audience with high need of information , convenience and entertainment tend to use Internet much more longer, which means the exposure to advertising is increased and result in the indirectly improvement of the effect of the interactive advertising. A great deal of research of effect mechanism has based on a core variable, that is, interactivity. There are lots of effect model put by scholars. The Integrative processing model of interactive advertising (Shelly and Esther2000) is a typical one of them (please see figure1). The basic conception of the model is that there are two parts of the information processing under the interactive circumstance, one is the function, the other is structure. Each individual was the starting point of the information processing episode in the context of interactive, and they would adapt themselves to different interactive context to satisfy their needs and goals. In this process, individual is the active sponsors and involvers. Additionally, Shelly and Esther's model suggested an integrative and two-way process which embedded the consumer-controlled factors and the advertisers-controlled factors together, however, their notion of this model still has some limitations with no micro-level analysis on interactivity effectiveness mechanism of generate and impact paradigms.

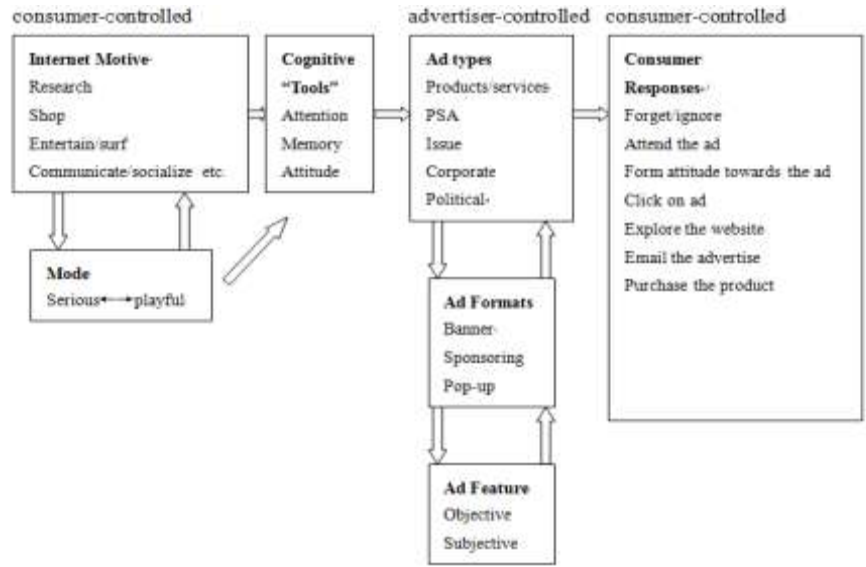

Figure1. Integrative processing model of interactive advertising

Thus, Wendy (2004) has examined how interactivity influences the ads effectiveness among consumers and what is the relationship among interactivity, product involvement and web experiences in micro-level. The Path Diagram of the Structural Equation Model proposed by Wendy Macias (please see figure2) convinced that interactivity has direct impact on enhancing the advertising comprehension and persuasion. The overall assumption of the model is that the interactivity is an important factor in both the consumer's comprehension of interactive advertising and the persuasive outcomes (attitudes and purchase intention). Furthermore, higher product involvement, higher understanding of the ads and the website, also web experiences and the advertising comprehension and persuasion are positive related. In a word, interactive ads can help improve the understanding of advertising information with the characteristics of persuasive.

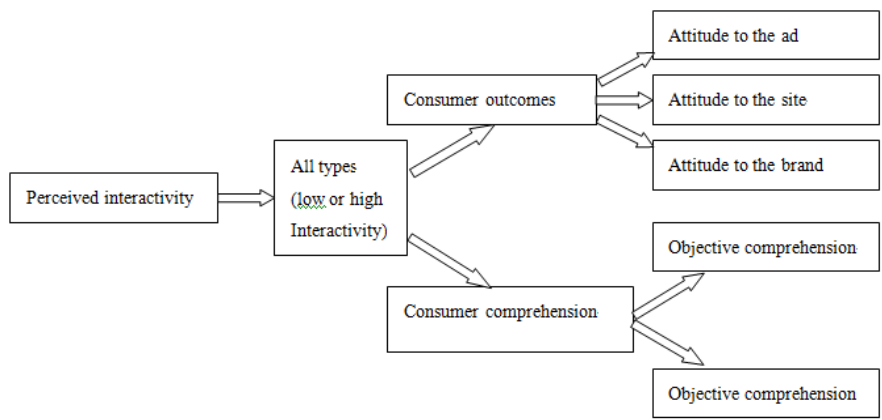

Figure2. The Path Diagram of the Structural Equation Model

Additionally, a Dual-Process Model of Interactivity Effects pointed by Yuping Liu and Shrum (2009)(please see figure3),with a new perspective which is proposed and tested that posits differential effects of interactivity on persuasion depending on person and situation factors, also it provides a three-dimension model which explain the relationship among interactivity, involvement and experiences. The central tenet of the model is that interactivity can affect the effectiveness of persuasion through distinctive processes, either by severing as peripheral cue through its mere presence in a website or by directly interacting with central processing (through the facilitating and/or inhabiting effects and through interactivity serving as a central argument itself). Audience with high involvement and web experience tend to using interactive options much more often and in much deeper level, which is benefit to improve the favorable impression to brand, while audience with high involvement but low web experience tends to using interactive options badly, which will decrease the favorable impression to brand. Generally speaking, Yuping's model has made some breakthrough compared to former research, put another way, not only did it examined the positive effect of IA, but it also suggested negative effect through the dual-process.

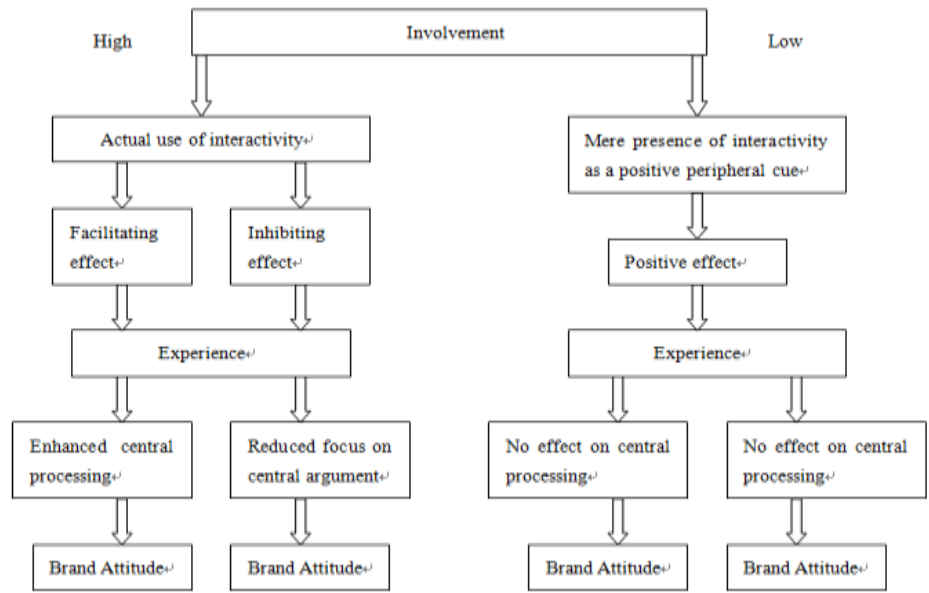

Figure3. Dual-Process Model of Interactivity Effects

In sum, the IA inner communication mechanism research has undergone a modifying trend gradually, from the macro perspective to the micro perspective. Integrative Processing Model of Interactive advertising with the biggest advantage of the combination of traditional advertising model and interactive advertising model, as well as many other factors related to advertising effect research into one model such as ads types, formation, consumer cognition, attitude, memory and so on, while this model was lacking of specific analysis of the interactivity's generate and influence path in micro level. Then the Structural Equation Model proposed by Wendy Micias(2004) modified it regard the specific and micro analysis, which pointed out the interactivity's core impact and how it works by the paired samples T-test of core interactivity. Finally, the Dual-Process Model of Interactivity Effects has further innovation compared to Wendy's model, firstly, it focused on the degree of audience's participation and the website experience rather than interactivity which embraced audience-centered value, secondly, it broke the research limitation focused on positive effect only ,by reveal the negative effect.

However, these three models above has limitations with lacking of the clear description and analysis of the communication levels and spread path from micro to macro for interactive advertising, which is the research purpose of this article.

## 3. THE THREE CASES AND ANALYSIS

In recent years, there have been several incidents in which the research of IA's effectiveness mechanism has focused on interactivity, while just with different emphases. In this section, the author focused on three types of IA that with unique characteristics to determine a effect model include different levels, types of the interacting episode in advertising though the case analysis method. Meanwhile, the effect model in this article highlighted the 'self' function during this process, that is to say, enhancing the effectiveness of advertising communication through some inner factors such as the options of interactivity and the dimensions of interactivity.

## 3.1. Official website interactive Ads

Official website, the concentrated expression of the brand image, provides the direct channel in B2C communication process. In the new media era, the audience-centered view dominants the market gradually. The traditional official website which provide the basic introduction for company, products and services is not enough to satisfy the consumer's need, furthermore, consumer asked for more functions such as a communication platform connected consumer and corporation, consumer and product, consumer and consumer, which is required by the interactivity characteristic of the Internet. In recent days, almost every brand official website provides some interactivity options or features, such as clicking service, comments, purchasing links and so on. But how dose these interactivity options and features work and what kinds of communication effect did they have, author applied the recruitment ads in the official website of the Tourism Queensland of Australia in 2009 as a typical case to analyze.
( www.islandreefjob.com )

The Tourism Queensland of Australia designed and posted e recruitment ads on its' official website with the topic of 'the best job around the world', that is, the caretaker of the Great Barrier Reef. Tourism asked all applicants download the application form and completed it on official website as well as they were required to make a candidate video which less than 60secs, and uploaded it to YOUTUBE, and finally, officials selected a winner through online voting as well as other election approaches. Honestly speaking, this recruitment ad is more like a tourism promotion ads, which based on the rich resources of official website, therefore, they attracted numerous tourists to visit their ads website and aroused wide attendance, which is more effective compared to the traditional recruitment ads. According to the official statistics provided by the Tourism Queensland of Australia, they have poured into large amount of money which up to more than one million U.S. dollars, and in result, they have achieved more than 70 million PR value. Also, this interactive ad has been granted several international rewards such as Creo Interactive Media Award, the gold medal of the digital interactive media in New York Advertising Festival and so on, and this is one of the reasons why author chose it as the typical case to analyze.

Throughout the whole design of the interactive advertising, the author summarized it form four aspects by case study, that is to say, these four aspects were key points boost the huge success. Please see figure4 as the basic process of this case for the core mechanism model of official website interactive ads.

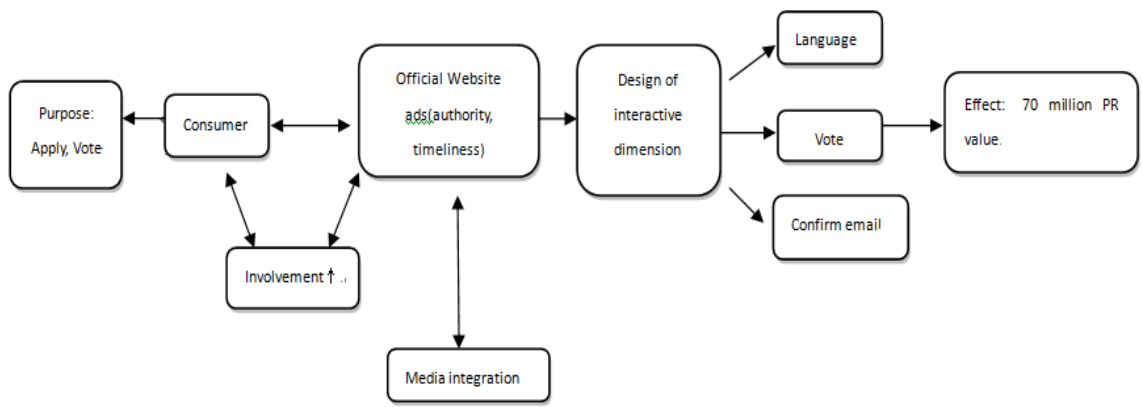

Figure4.Core mechanism of Official Website IA

### 3.1.1. Interactive options and active type enhancement.

Overall, all the interaction options on the official website of the Queensland Tourism promoted the public participation and achieved an active type effect enhancement in result. Firstly, they have set up five different home page versions in view of different languages, which included English version, Chinese version (Simplified and Traditional), Japanese version, Korean version and German version. There is no doubt that the language conversion option granted the maximum participation due to the borderless of Internet. Secondly, they design an online voting option, during voting process, not only the applicants would click and browse the interactive ads, but voters, they would share the applicants videos, exchange their points and vote, all of these activities are conducive to the independent participation and active interaction, and more importantly, during this kind of process, the participants included applicants and voters would get more information about Great Barrier Reef and deeper understanding of this advertising as well as their tourism program. And more frequently they interactive with each other, much deeper they understand and memorize about these 'inner' information. Finally, the Email confirmation option designed by officials ensured that the Queensland Tourism could send their pushing advertising to the final audience directly. During the whole interactive advertising, voters were asked for providing their Email address, only providing it they can vote online, then they will received a conformation Email from Queensland Tourism, after confirming this Email, they can vote finally. During the conforming process, every voter would browse the beautiful pictures of Great Barrier Reef bundling with Email unavoidably, and also, they would receive Greeting Emails from Great Barrier Reef from time to time, therefore, participants would influenced by the tourism information imperceptibly. Sum up, this advertising contained three interactive dimensions which are, the human-information interaction (the tourism pictures), the human-human interaction (the video share between applicants and voters) and the human-computer interaction (all kinds of interactive options), participants would enhance their understanding and memory of the Great Barrier Reef, and we can image that after this advertising, the participants maybe have urge to travel there, in this way, the effect of official website interactive advertising has achieved self-enhanced by interaction gradually.

### 3.1.2. The media integration, interaction and UGC.

In this case, author found that the media integration and interaction also make great contribution to the advertising effect's self-enhanced form the perception of UGC, which means user generate content. On the one hand, the Queensland Tourism officials advertised on their official website, which granted authority, also they integrate official website and YOUTUBE, blog together during the voting process, on the other hand, they achieved viral diffusion of advertising information which is very fast through the integration and interaction between different media in a broader level. For instance, lots of participants logged in major campus forum, major online community such as Tianya and post, the participants interact with others to share as well as generate new points and new information about this ads by the copy, paste, send, share and other kinds of online interaction. There are statistics show that if you search pictures by the title of 'the best job around the world' in Flicker, you will get 4484 results, and there are over 578hours et al video length related to the advertising on YOUTUBE . Therefore, the official website advertising ensured that the tourism information penetrate among cyber citizens and improve the advertising effect, through the integration with other media especially with social media, and the interaction between different media in macro level.

### 3.1.3. The agenda-setting and sensational advertising effect.

It is very obvious that new media and online media agenda interact together to promote the advertising effect in this case. Firstly, online media disseminated the recruitment ads over the Internet widely to set up a hot topic, then the traditional media join in to set agenda in company with new media. New media let topic disseminated rapidly with the characteristics of quick and short cycle, while traditional media granted higher authority with the characteristic of orthodox. Therefore, new media and traditional media interact with each other to settle an interactive and circular network to 'disseminate-and-deepen', resulting in a snowballing effect of advertising communication.

### 3.1.4. The timeliness of official website.

Also, the officials achieved self-enhanced effect of IA by taking advantage of the timeliness of official website. Marshall Mcluhan foresaw and pointed out the concept of 'global village' last century, he suggested that people around the world could focus on same thing at the same time regardless of the time differences and geographical limitation due to the timeliness of Internet.

Apply this theory into the interactive area of Internet, we can see that the timeliness of official website interactive advertising is incomparable huge compared with other kinds of interactive advertising, owning to the information updated timely. In this case, Queensland Tourism updated related information in time according to the advertising process, which affording participants a platform to simultaneous feedback as well as an interactive space to interact timely. That is to say, the timeliness of official website guarantee participants and more cyber citizens can interact in time, also guarantee the timeliness of self-enhanced effect.

**3.2. SNS Interactive Ads**

SNS, abbreviated from a name of 'social networking services', which is a product mix for social media or social networking combined with human communication and mass communication in new media era. Author selected 'Lohas(COFCO)'and 'lucky line(UNIQLO )  as typical cases in this part and whole analysis was focused on the advertising self-enhanced effect based on three main aspects. See figure5 as the basic process of this case for the core mechanism model of SNS interactive ads. Please see picture1 and picture 2 to identify the homepage of the interactive advertising of Lohas and Uniqlo.

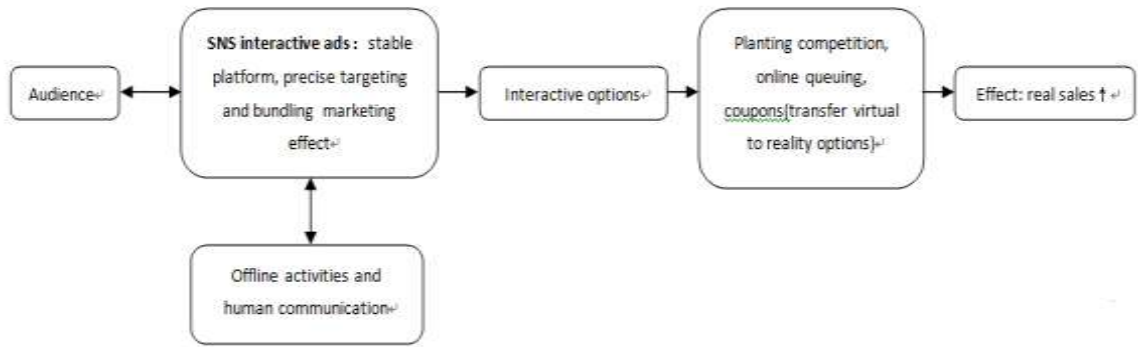

Figure5.Core mechanism of SNS IA

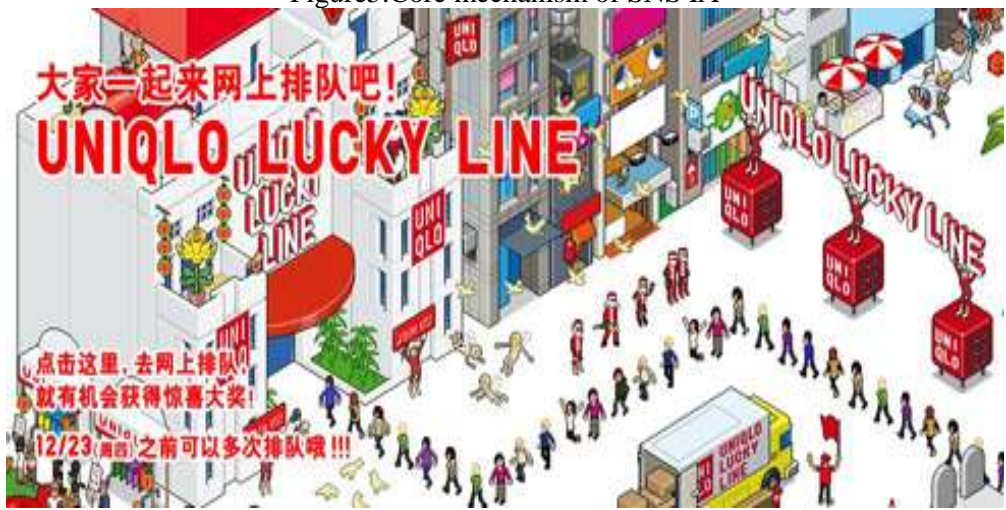

Pic.1.The screen capture of Uniqlo's lucky line

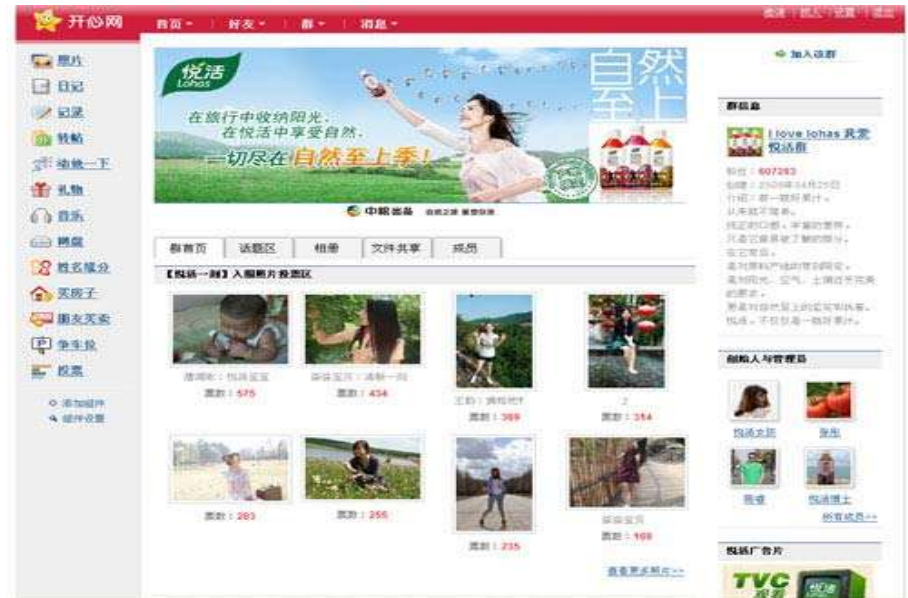

Pic.2.The screen capture of Lohas IA on Happy net

### 3.2.1. Unique interactive options.

From the two cases above, author found that there are some unique design of interactive options which was character with SNS closely, an Internet appliance service to help cyber citizens to settle social service networking and social platform. First of all, they made precise ads targeting by using proper social media as their marketing position. Lohas cooperated with Happy net (a social network in China), and UNIQLO made cooperation with Renren (a famous social network among Chinese youth), both Lohas and UNIQLO are targeting on young consumer, choosing an appropriate advertising media is significant premise to guarantee interact smoothly. Next, users-bundled marketing based on SNS is a stable platform for frequent interactions. For example, Lohas cooperated with Happy net mainly focused on an interactive advertising marketing named 'Lohas Planting Competition' which was last more than a month, participants would go through there stages which are 'natural first experience', 'natural vitality season' and 'natural supremacy season'. Also participants were asked to binding their Happy net account to simulate planting activities such as sowing, watering, fertilizing, reaping juicing, bargaining and so on, which means every participants would interact with Lohas drinking product or even the whole interactive advertising more than just once, they must interact several times and repeat planting activities many times to win the competition, in that way, participant would processed a deeper memory and understanding for Lohas product advertised or even Lohas brand unavoidably, and statistics show that the actual sales of Lohas drinking has improve 30 percentage within one month the IA launched. Finally, 'Synchronized to account' option attracted more audience to attend interaction, in the case of UNIQLO 'lucky line', there was an option to update their Renren account status by share 'lucky line' activities, which boost a huge participation to this interactive advertising marketing, furthermore, not only can participants interact with the UNIQLO products, but they also can talk with friends online which is human-human interaction. In total , there was 1,330,496 people have attended this interactive advertising and the number of people participant has break 100,000 only on the first day of activity.

### 3.2.2. Online and offline interact.

Combine online interaction and offline interaction together is an immediate and effective way to realize self-enhanced effect of interactive advertising from the analysis for the chosen two cases. UNIQLO provided every 'lucky line' participant coupon online for offline shopping, that is to say, participants could enjoy discounts in UNIQLO stores by using these coupons within the specified time. Similarly, Lohas held 'Lohas Planting Competition' online for coordinating direct sale offline. For example, Lohas once held an interactive game named 'Lohas Origin

Tour' in JOY CITY of XIDAN in BEIJING, the whole game is connected closely with Lohas' interactive ads online, embedded all kinds of brand information of Lohas drinking come from their online ads, and if participants understood and memorized product and brand information enough, they would win the game and reward a free tour to the origin place of Lohas. And statistics show that the actual sales of Lohas drinking have improved 30 percentages within one month the IA launched without other kinds of advertisements promotion.

### 3.2.3. Human communication and mass communication.

Finally, human communication process and mass communication process was interacted with each other during the whole interactive advertising to self-enhance advertisement effect. Actually, SNS itself is the outcome that human communication merge in mass communication in new media era, thus, interactive advertising which based on SNS is another platform to combine human communication and mass communication unavoidably. In our cases, human communication brought out 'Multiply Effect' for advertising effect by broadening consumer participation and media agenda-setting, put it in another way, human communication help to set a positive interactive chain from consumer participate –social media –opinion leader – more consumer participation. In SNS, even strangers can establish contact with each other quickly and follow media hot topics, SNS with the characteristics with low accessibility and openness make opinion leaders can deliver their viewpoints with others conveniently, also opinion leaders can share interests and desires with others to created a communication exciting point to convey the advertising information to consumers from interaction of the level of communication forms.

### 3.3. Mobile Media Interactive Ads

Mobile media includes individual media represented by mobile phone, pad and public media with representative of car terminal. Author convinced that technology-oriented is an obvious feature during the history of mobile media's development, that is to say, every advent of new technology accompanied with a new marketing opportunities, such as new APPs, LBS (location bases services), VR ( virtual reality) and so on.

There are limited related researches concerned on this area, so author selected two typical cases on individual and public area separately and analyzed effect mechanism. Durex developed a game named 'babe plan' in APPLE APP STORE to connect Durex product with iphone users, put it in another way, 'babe plan' combined the carrier (mobile media)and the core content( interactive advertising ) together organically to achieve self-enhanced effect. Specifically speaking, there are strong correlation of target market between iphone and Durex, which both are targeted on passionate, energetic young people who pursuing fashion and trends, using iphone as well as its' mobile game can implement product placement invisibly and maximize the contact with target audience. In other place, the interactivity embedded in this game provided users a chance to interact with Durex brand information intentionally or unintentionally, that is to say, Durex brand image impressed them more or less intentionally or unintentionally and Durex can establish better emotional connection with consumers. Also, Durex made video about the operation of this game app and uploaded it to main website globally, such kind of "virus-video" communication model increased the download amount of the game as well as improved the game's influence. In Youku, there was about 9,400,000 traffic for the video called " iphone app for Durex Babe Plan".

Similarly, Subway once promoted sandwiches by car media terminal, they designed a car media terminal 'making sandwiches' game on taxi, passengers can simulate sandwiches making process to make an sandwiches while taking taxi, while they finished making sandwiches, they would rewarded a coupon and a chance to have discount. There are statistics shows that more than 5 million people have watched this advertisement and 70,000 people have clicked and participated in this interactive advertising and 10433 people have sent short message to get electronic discount coupons only sampled with 4000 vehicles in Shanghai within one month.

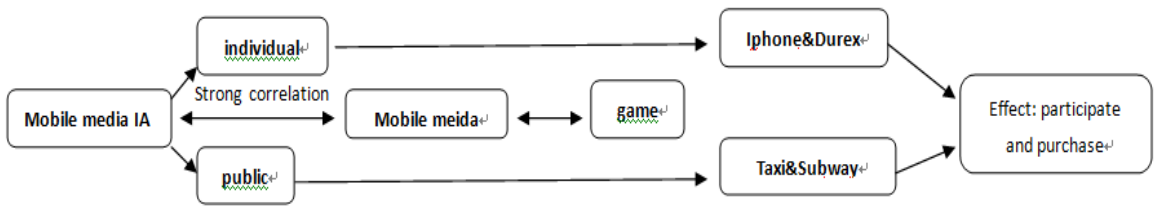

Figure6.Core mechanism of mobile media IA

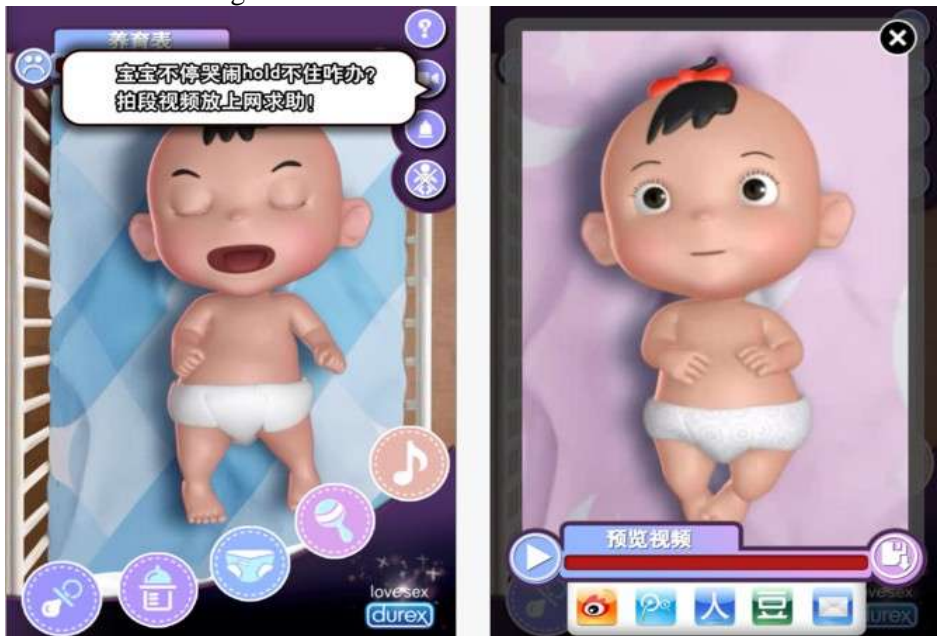

Pic.3.The screen capture of Durex's babe plan

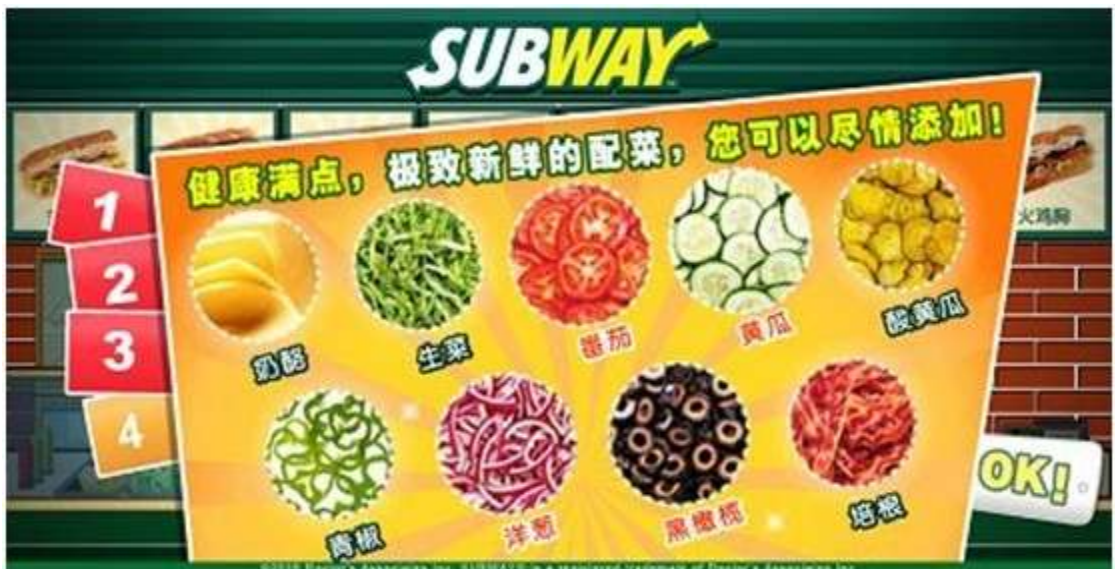

Pic.4.The interactive game interface of Subway in taxi

# 4. RESULT AND DISCUSSION

These three case studies appear to show that the self-enhanced mechanism of interactive advertising which can be specified into three parts, spiraling increasing trend in macro level, fission type communication approach in meso level, and the specific interactive levels and process in micro level.

In the first place, figure7 shows the micro level mechanism, there are three levels interaction which are core interaction, inner interaction and outer interaction to achieve the self-enhanced effect. Specifically speaking, core interaction level also the dimensions of interactivity, that is human-competition, human-content interaction, inner interaction level refers to the interaction with advertisement and the improvement of the advertising cognition and understanding, while outer interaction level means the interaction with other media and the whole process of integrated communication and marketing.

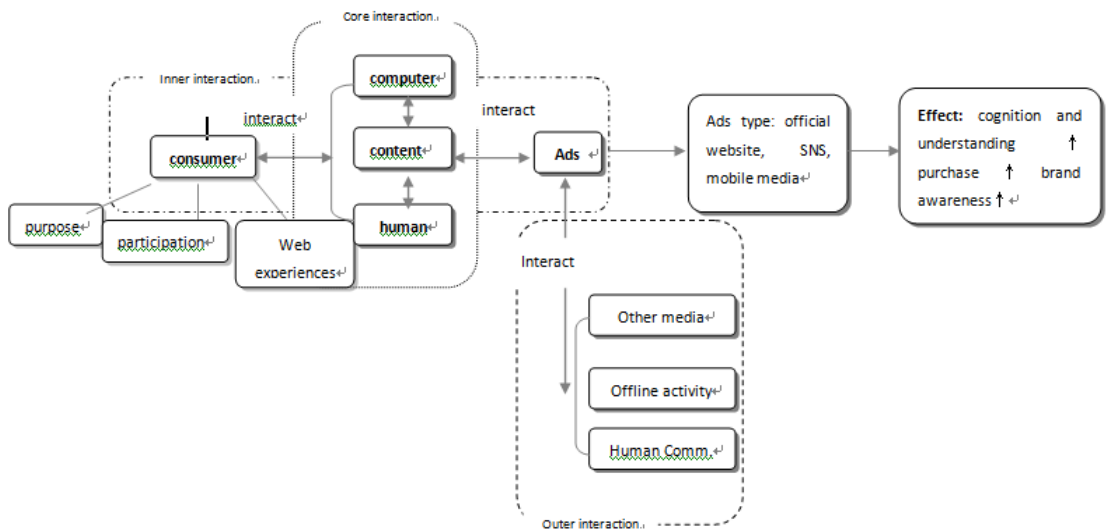

Figure7.Micro level mechanism of self-enhanced model

In the second place, fission type communication merged by the interaction in meso level shown in figure8, the advertising information passed from one to two, two to four, four to eight, and the advertising information underwent a round-trip in the interaction platform, thus the cognition and understanding of advertising information achieved self-fission-type spread and self-enhance.

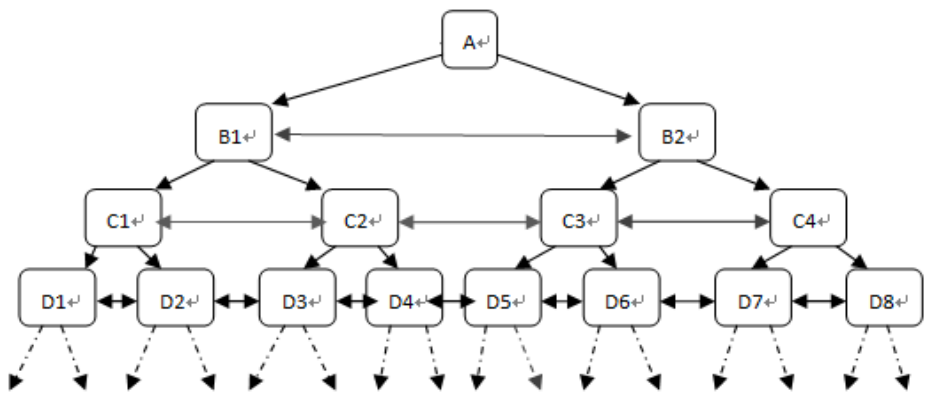

Figure8. Meso level mechanism of self-enhanced model

In the third place, figure 8 shows the macro level effect of self-enhanced model which is similar with the Dance spiral model and Spiral of Silence. In the long term, the communication effect of interactive advertising is rising cumulatively generally. In this article, interactive advertising brought out huge PR values, gained wide-range public attention or promoted sale amount of product and brand awareness. In a word, interactive advertising realized the spiral increasing effect by different levels interaction.

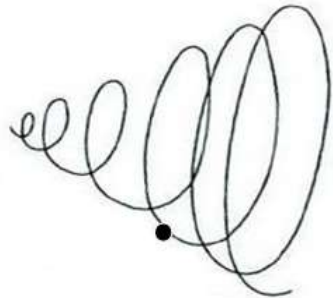

Figure9.Macro level level mechanism of self-enhanced model

Author combine these three parts together shown in figure 9, there is a core during the whole process which is interactivity, each type interactive advertising chosen in this article connected closely with interactive options includes human-computer interactive options, human-content and human-human interactive options.

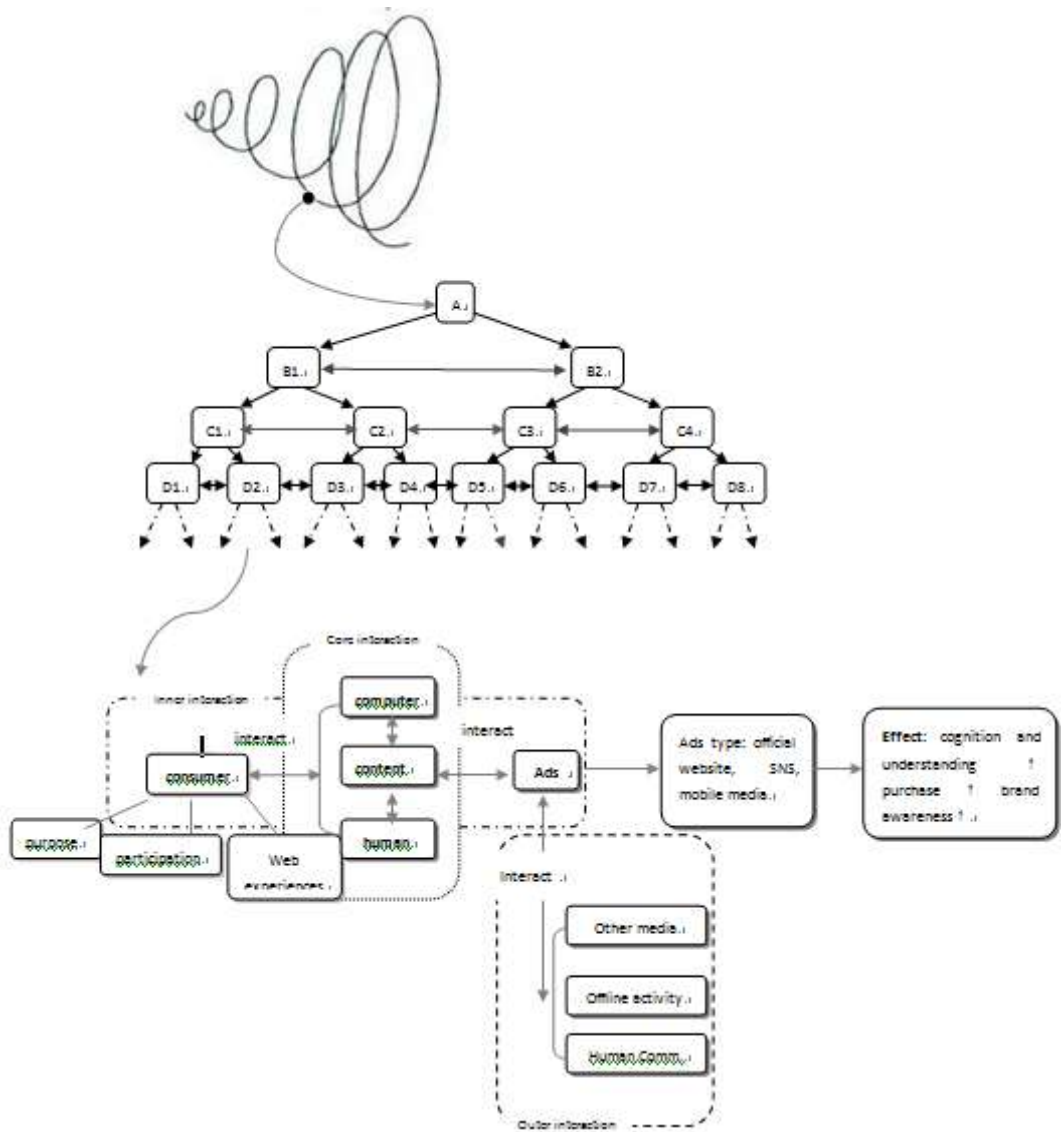

Figure10.Self-enhanced model of IA

Last but not least, the core concept of self-enhanced model is 'self'. On the one hand, the higher degree of interactivity, the more likely to realize sum non-zero effect which is also known as win-win. High degree of interactivity an advertisement possessed means more levels and types connection and a more multi-dimensional and interdependence network which is propitious to strengthen the relationship between advertisers and consumers. Such kind of multi-dimensional interdependent network granted the self-enhanced and mutually beneficial relationship between consumer and advertiser. For example, in the case of UNIQLO's 'lucky line', when participants got real coupon by attending the online interactive advertising, they will have more wishes to attend this advertising for more times to get more coupons, such kind of interaction makes the model of "more three wishes" came true. On the other hand, the more times to interaction, the better self-enhanced effect. Kevin Kelly suggested that cumulative society organization showed some pure mathematic things beyond good-neighborly relations, and the increment system is based on information flow which is crucial to format interdependence network to connect trust and competition together, also while the links within the network increased, the increasing and speeding power is enlarged. Take this assumption into interactive advertising, the increment system of self-enhanced model is based on the information flow granted by interactive options, that is to say, more links, options of interactivity, deeper interaction consumer participated and more powerful self-enhanced effect.

## 5. QUESTION FOR FURTHER RESEARCH

This article analyzed three typical cases of the appliance of interactive advertising and pointed out self-enhanced effect model creatively based on the review of interactivity and related researches. However, this article is lacking empirical and quantitative research, and with the limitation of small samples.

Three questions for further research follow from our study, first one is the quantitative analysis related to this model, the second question for the larger samples, finally, further research should consider a wider array of interactivity characteristics and levels to identify the effects of different interactive tactics.

## ACKNOWLEDGEMENT

This article was supported by China National Social Science Foundation in 2014and the Hujiang Foundation of China.

**Authors**

Renjian, **Tel. +86 136 0182 8470**

**Add. No.516Jungong Rd. Yangpu District, Shanghai, China. 200093**

professor, College of Communication and Art Design, University of Shanghai for Science and Technology. Ph.D of Economics, Fudan University. Postdoctoral of Management, Tongji University. Mainly focus on the research of brand communication and new media communication.

Renjian

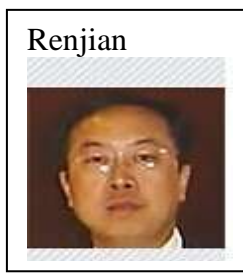

Ding Wanxing, **Tel. +86 183 1715 9290**

**Add. No.516Jungong Rd. Yangpu District, Shanghai, China. 200093**

Master, College of Communication and Art Design, University for Shanghai of Science and Technology. The main research direction is interactive communication and interactive advertising.

DingWanxing

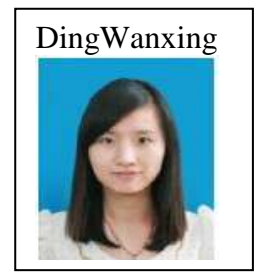